\documentclass[pra,twocolumn,superscriptaddress,floatfix,amsmath,nofootinbib,amssymb]{revtex4-1}

\usepackage{graphicx}
\usepackage{bm}
\usepackage{verbatim}
\usepackage{mathrsfs}
\usepackage{color}
\usepackage{paralist}
\usepackage{hyperref}
\usepackage{subfigure}

\newcommand{\ii}{\mathrm{i}}

\newcommand{\ket}[1]{| {#1} \rangle}

\begin{document}

\title{ Mode-Invisibility as a  non-Destructive Probe of Entangled QUBIT-CAT States}
\author{Paulina Corona-Ugalde}
\email{pcoronau@uwaterloo.ca}
\affiliation{Institute for Quantum Computing, University of Waterloo, Waterloo, Ontario, N2L 3G1, Canada}
\affiliation{Department of Physics \& Astronomy, University of Waterloo,  Ontario Canada N2L 3G1}
\author{Marvellous Onuma-Kalu}
\email{monumaka@uwaterloo.ca}
\affiliation{Department of Physics \& Astronomy, University of Waterloo,  Ontario Canada N2L 3G1}
\author{Robert B. Mann}
\email{rbmann@uwaterloo.ca}
\affiliation{Department of Physics \& Astronomy, University of Waterloo,  Ontario Canada N2L 3G1}

\begin{abstract}
 We investigate the dynamics for a two level atomic system entangled to coherent states using the recently developed
mode invisibility  technique.  Using a quantum 2-level probe,  we demonstrate a way to non-destructively measure a number of properties between a qubit  entangled with a generalized CAT state, including the amplitude of the coherent state, the location and relative excitation of the qubit,
and the von Neumann entropy.  Our results indicate a connection between this last quantity and the interferometric phase shift of the probe, thereby suggesting a possible way to  experimentally measure entanglement non-destructively.
\end{abstract}

\maketitle

\section{Introduction}

Quantum mechanics admits a host of states, such as those involving the coherent superposition of two or more macroscopically separated localized states that do not exist in classical physics.   This perplexing situation has troubled scientists 
since the earliest days of quantum physics, and for decades remained out of the reach of experimentalists to actively probe.  However over the past decade or so it became possible to construct quantum superpositions of mesoscopic field states  trapped in one or more cavities, rendering this  phenomenon open to empirical study \cite{Davidovich}.   Several models have been proposed  \cite{HWG,Haroche,Davidovich,cavities1} as to how to both generate and detect  so-called `cat' states, namely a superposition with equal probabilities of two coherent states with different complex amplitudes.  In one of these proposals  \cite{Davidovich}, two single two-level atoms  (whose states are $\ket{g}, \ket{e}$) were used, one for creating the cat state and the other for detecting the interference. For carefully selected atomic velocity and frequency, the first atom in the state $(\ket{e} + \ket{g})/\sqrt{2}$, was made to interact non-resonantly with a coherent field $\ket{\alpha}$  \cite{coinco} suspended in a superconducting cavity,  thereby forming the combined state 
\begin{align}\label{sstate}
 \frac{1}{\sqrt{2}}\Big(\ket{g,\alpha} + \ket{e,-\alpha} \Big)
\end{align}
called a Bell-Cat state \cite{nature2015}.

The entangled state \eqref{sstate} can be probed and/or manipulated by sending a second atom into the cavity. In the scheme discussed above \cite{Davidovich},  it is possible to generate a cat state $1/\sqrt{2}(\ket{\alpha} \pm \ket{-\alpha})$ by subjecting the atomic states $\ket{e}, \ket{g}$ to a desirable frequency. A cat state is easily distinguishable from a statistical mixture via its coherence terms. However these coherence terms are easily lost to the environment and one observes a decay of the cat state to a statistical mixture. Indeed, the coherent control of such mesoscopic systems, and harnessing their quantum  interference or entanglement against the detrimental effects of the environment is a major  issue \cite{prl70,Davidovich,OpenQM} with implications not only for the foundations of quantum mechanics but also for  quantum computing, quantum simulations of many-body systems, cryptography, and quantum metrology.

 In this paper we consider a more general and complex quantum state of the form
\begin{align}\label{estate}
\ket{\psi} = A\ket{g, \alpha} + B\ket{e, \beta},
\end{align}
produced in the cavity; $A$ and $B$ are real coefficients that characterize the atom-field entanglement while $\ket{\alpha}$ and $\ket{\beta}$ are arbitrary coherent states. We do not subject the atom in this entangled state to further measurement but rather send in a second atom to probe the entangled atom-field system. 

In this setting, one can ask a broader range of questions. Can one experimentally detect  and quantify the atom-field entanglement in a non-destructive way?  How much information about the joint state do we have?  What is the entropy of entanglement associated with this system? A key aim of this paper  is to provide answers to these questions.

Different methods have been proposed for detecting entanglement \cite{QM,nature,nature2015,Emeasures,Moura2004}.  The most popular approach to the problem of  detecting the presence of entanglement in ``Bell-cat states'' is based on violation of Bell-type inequalities \cite{nature2015}. One can also perform complete quantum-state tomography  \cite{holevo}. However these methods are not only a destructive process, but also  pose  both fundamental and  technical challenges.  Another method entails measuring entanglement witnesses. This is efficient, but not universal and requires information about the state prior to its measurement \cite{pra2000,Terhal2000}.

Both theory and experiment  have progressed to a point where, by trapping electromagnetic (EM) fields in a highly reflective cavity and using appropriate detectors as probes in an interferometric setting, many quantum non-demolition measurements  can be achieved to considerable precision \cite{Brune1996,serge,nature,cavities1}.  To define a quantum non-demolition (QND) measurement scheme \cite{qndhistory}, physical states of the EM field are left significantly unaffected after the measurement process and information acquired from the state is large enough to enable  repetitive measurements \cite{serge,marvy2013,marvy2014}. 

We  consider here a  scheme that detects the general state of the form in \eqref{estate}. Our aim is to acquire significant information from this  detection whilst negligibly perturbing the entangled atom-field state. Specifically we shall consider the mode-invisibility measurement technique \cite{marvy2013,marvy2014} -- a QND measurement scheme with several advantages.   It leaves a given quantum system almost completely unperturbed after measurement. Measurement can therefore be repeated on the same quantum system since a first measurement leaves the system's state significantly unperturbed, and so can yield a large amount of information about a system.

We will investigate the applicability of the mode invisibility scheme to measure the relevant features of the entangled state \eqref{estate}. We assume that we have no information about the state's entanglement or alternatively partial information about the features (amplitude and phase) of the coherent field states $\ket{\alpha}$ and $\ket{\beta}$ and  will investigate to what extent we can obtain this information without significantly perturbing the quantum state. 
 
  We begin in section \ref{review} with a brief review of the mode-invisibility measurement scheme \cite{marvy2013}. To obtain information  about the entangled state, we need to suspend it in a single mode of a highly reflective cavity \cite{Haroche} and use  a detector to probe the occupied cavity mode. This detector-cavity mode interaction and various measurement criteria for the cavity mode is dicussed in section \ref{twos}.  A detector that interacts with   the field in the cavity will, via the mode invisibility technique,  acquire a global phase factor that carries information about the cavity content. To measure this phase factor, one needs to define an appropriate reference state.  In section \ref{three}, we consider  how the global phase from the detector-cavity mode interaction can be measured and we also characterize the information content of the constituents of the entangled state \eqref{estate}.  We consider in section \ref{six}  the phase resolution and  interferometric visibility under the condition that the measurement process is non-destructive.
  We summarize our results in a concluding section.

\section{Brief review of the mode-invisibility measurement scheme}\label{review}

Here we give a review of the mode invisibility technique (for a detailed explanation, see \cite{marvy2013}) . The measurement setup is an atom interferometer (a type of Mach-Zehnder interferometer) which consists of two highly reflective optical cavities of length $L$, each placed along the distinct branches of the atomic interferometer.  An unknown field state (whose properties we want to measure) is stored in one of the cavities, while a reference quantum system (whose properties we know)  is stored in the second cavity. To probe the unknown state we employ a detector (or probe) modelled as a two-level Unruh de Witt detector\cite{UnruhB,Unruh,Quantumthermometer}. The detector enters  the atom interferometer at  constant velocity $v = L/t$ , where $t$ is the interaction time.   As it makes its way into the atomic interferometer, it splits into a superposition of two partial beams, each part  propagating along the branches of the interferometer and interacting resonantly with the respective cavity contents along the super-posed trajectories. 

When a detector interacts on resonance with the cavity mode, it does so strongly, in general altering the field state\cite{sculbuk}. Such an interaction enables one to gain significant information about the field (in a case where the atom is used as a detector to probe a field trapped in a cavity). On the other hand, a strong interaction leads to alteration of the cavity mode thereby jeopardizing the QND measurement criteria. One needs therefore a scenario where the alteration in the cavity mode is minimized whilst permitting acquistion of information. 
\begin{figure}
\begin{center}
\includegraphics[width=.25\textwidth]{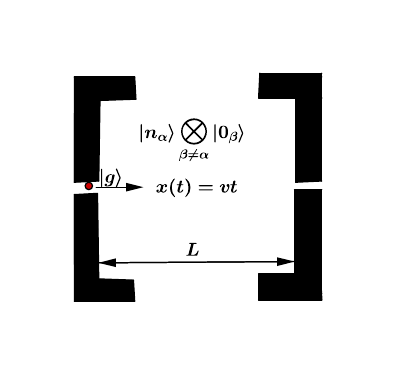}\includegraphics[width=.25\textwidth]{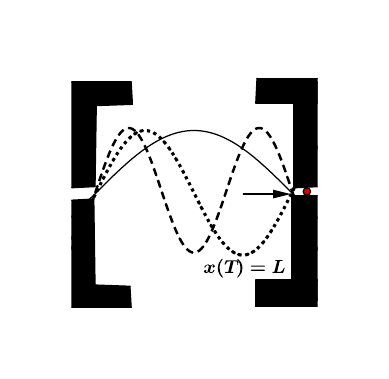}
\caption{(Color online): Scheme of an atom in its ground state $|g\rangle$ going through a cavity at constant speed. The cavity modes are in a state $ \ket{n_\alpha}\bigotimes_{\beta\neq \alpha}\ket{0_\beta}$.  Atoms resonant with even modes preserve our approximation \eqref{phasee} and   acquire a significant global phase.}
\label{waves}
\end{center}
\end{figure}

 Mode-invisibility exploits the fact that the spatial symmetry of the even field modes in a cavity effectively cancels the action that an atomic probe transversing the cavity at constant speed exerts on the field. This is illustrated in figure \ref{waves}, which shows the motion of the probe qubit through the cavity. When the probe  interacts only with the even modes of the cavity ( dotted line ),  the changes caused by its interaction during the first half of its motion through the cavity $x\in[0,L/2]$ are  undone as   it travels through the second half of its motion $x\in[L/2,L]$ and exits the cavity \cite{Dicke55,Stephen98, Sarkisyan04}. This is possible because the effective sign of the coupling to the cavity ($\lambda$ times the spatial distribution of the mode) reverses half way through the flight path of the atom. This will have the advantage that we can cancel out the leading order contribution to the transition amplitude for the field and the detector while keeping constant the leading order in the phase shift.


\section{Mode Invisibility as a Probe of Entangled Qubit-Cat States}\label{twos}

We next turn our attention to entangled qubit-cat state measurements.  Our interest is in characterizing the entangled state  \eqref{estate}.  In other words we want to (non-destructively) acquire information about the parameters $A, B, \alpha$ and $\beta$, where $\alpha = |\alpha|e^{\ii \theta} $ and $\beta = |\beta|e^{\ii \phi}$ defines the complex amplitude of the coherent field states $\lbrace\ket{\alpha}, \ket{\beta}\rbrace$ respectively.   We shall choose $\theta=-\phi$ throughout in order to
obtain a Cat state in the limit $|\beta|\to |\alpha|$.
As discussed earlier, the probe is modelled as 
 an Unruh de-Witt detector \cite{Unruh} prepared initially  in its ground state so that  at time $t=0$ 
\begin{align}\label{joint}
\ket{\psi(0)}=\ket{g_{p}}\Big(A\ket{g_{q}}\ket{\alpha_{\kappa} }+ B\ket{e_{q}} \ket{\beta_{\kappa}}\Big)\bigotimes_{\kappa \neq \varsigma}\ket{0_\varsigma}.
\end{align}
The subscript $q$ denotes the qubit that is entangled with the field state; we refer to this as the cavity-qubit so as to   distinguish it from the state of our probe which will be given the subscript $p$. We have assumed that the coherent field states $\ket{\alpha_{\kappa}}$ and $\ket{\beta_{\kappa}}$ are trapped in the cavity mode $\kappa$ having frequency  $\omega_{\kappa}= c k_{\kappa }$ with $k_{\kappa} = \kappa \pi/L$ being the wave number, while the rest of the cavity modes are less occupied (approximately vacuum).   Here $c$ is the speed of light.

Now we want to estimate the dynamics of the combined system from a given time $t=0$ to $t=T>0$ where $T = L/v$ is the interaction time.  In the interaction picture,
 $U(T,0) = \mathcal{T}\text{exp}\big[\frac{1}{\ii}\int_{0}^{T} dt H_{I}(t)\big]$ is the unitary evolution operator governing the 
 dynamics of the system; it  satisfies the properties $U(T,0)U^{\dagger}(T,0)=\openone$ with $U^{\dagger}(T,0) = U(0,T)$  so that at time $T$, the joint atom-entangled state is $\ket{\psi(T)} = U(0,T) \ket{\psi(0)}$. Expanding the time ordered exponential perturbatively, we obtain
\begin{equation*}
U(0,T) \!=\! \openone\!\underbrace{-\ii\!\!\int_{0}^{T}\!\!\!\!\!d t_1 H_{I}(t_1)}_{U^{(1)}}\underbrace{ -\!\!\int_{0}^{T}\!\!\!\!\!dt_1 \!\!\int_{0}^{t_1}\!\!\!\!\!dt_2\,    H_{I}(t_1) H_{I}(t_{2})}_{U^{(2)}}+\hdots
\end{equation*}
So that $\ket{\psi^{n}(T)} =U^{(n)}\ket{\psi(0)}$.  We assume that the qubits do not interact with each other but that the field undergoes an interaction with both, leaving us with an interaction Hamiltonian of the form
\begin{align}\label{hh}
\hat{H}_{I} = \lambda_{p} \hat{\mu}_{p}(t) \hat{\phi}[x_p(t)] +\lambda_{q} \hat{\mu}_{q}(t) \hat{\phi}[x_{q}(t)] 
\end{align}
where each term in (\ref{hh})  is  known as an Unruh de-Witt Hamiltonian \cite{Unruh}. The terms on the right hand side of (\ref{hh})   correspond to the  probe-field and qubit-field interactions, with  respective coupling strengths $\lambda_{p}$ and $\lambda_{q}$.  $\mu_{p}(t)$ and $\mu_{q}(t)$ are the monopole moments of the probe and qubit respectively.  We model our field system as a massless scalar field $\phi[x(t)]$, where $x(t) = vt$ describes the atomic trajectory through the cavity. For simplicity, we assume that the qubit entangled to a field state is at a fixed position $x_{0}$ in the cavity. For the detector's motion through the cavity, we consider Dirichlet boundary conditions (so that $\phi[0,t]=\phi[L,t] = 0$). In the interaction picture
\begin{align*}
\phi[x_{p}(t)] =  \sum_{\delta=1}^{\infty}( a_{\delta}^{\dagger} e^{\ii\omega_{\delta} t} + a_\delta e^{-\ii\omega_{\delta} t})  \frac{\sin[k_{\delta}x_p(t)] }{\sqrt{k_{\delta} L}}
\end{align*}
for the probe moving on a trajectory $x_{p}(t)$, whereas
\begin{align*}
\phi[x_{q}(t)] = \sum_{\delta=1}^{\infty}( a_{\delta}^{\dagger} e^{\ii\omega_{\delta} t} + a_\delta e^{-\ii\omega_{\delta} t})  \frac{\sin[k_{\delta}x_0]}{\sqrt{k_{\delta} L}}
\end{align*}
for the cavity qubit, and
\begin{align*}
\mu_{j}(t) = \ket{e_{j}}\langle g_{j}|e^{-\ii \Omega_{j} t } + \ket{g_{j}}\langle e_{j}|e^{\ii \Omega_{j} t}, \qquad j=p,q
\end{align*}
 where $\Omega_p$ and $\Omega_q$ are the respective probe and qubit frequencies. 
 Here $a^{\dagger}_{\kappa}$, $a_{\kappa}$ are the field creation and annihilation operators respectively obeying the commutation relation $[a_{\kappa},a^{\dagger}_{\lambda}] = \delta_{\kappa,\lambda}$. Since coherent states are eigenstates of the annihilation operator, the following relations
$$
\hat{a}_{\delta}\ket{\alpha_{\delta}} = \alpha_{\delta} \ket{\alpha_{\delta}} \qquad \langle \alpha_{\delta}| a^{\dagger}_{\delta} = \langle \alpha_{\delta}|\alpha^{*}_{\delta}$$
are valid.

 The mode invisibility technique assumes a resonant interaction between probe and cavity field, so that
$\Omega_{p} = \omega_{\kappa}$. We further assume  that the qubit's transition frequency is set off-resonance relative to the cavity mode frequency; in other words  $\Omega_{q} \neq \omega_{\kappa}$ with detuning $\delta = \omega_{\kappa}-\Omega_{q}$.

\subsection{Measurement Criteria}\label{mc}

Having presented our interaction model, we are now ready to describe the system's dynamics.   An external force applied to   two systems in an entangled state will in general modify the entanglement. As our goal is to non-destructively probe the entangled qubit-field system,   we wish to minimize (ideally eliminate) this effect.  This is the thrust of 
 the mode invisibility technique \cite{marvy2013,marvy2014}: it is designed to  ensure that a system undergoes an interaction with an external force in a non-destructive way.   
 
The measurement criteria for the mode invisibility technique are
\begin{enumerate}
\item The probabilities must be approximately unity \cite{EPR}
\begin{align}\label{hypo}
\big | \langle \psi (0) | U(0,T)| \psi(0) \rangle \big | ^{2} \simeq 1,
\end{align}
which means that the probability of transition of the probe into an excited state from its initial ground state is approximately zero. 
\item Provided that the criteria \eqref{hypo} is valid, the final state of the combined system (probe-entangled state) after the interaction time $T$ is approximately given by its initial state multiplied by a phase factor $\eta$. Mathematically, this implies
\begin{align}\label{phasee}
|\psi(T) \rangle = U(0,T)|\psi(0) \rangle \simeq  e^{i \eta}|\psi(0) \rangle  + \ket{\psi_{\perp}(T)}
\end{align}
\end{enumerate}
where $\langle \psi(0)|\psi_{\perp}(T)\rangle = 0$ and so 
\begin{align}\label{phased1}
\eta & =  - \ii \operatorname{ln}\Big[ \langle \psi(0)| U(0,T)|{\psi(0) }\rangle  \Big]
\end{align}
is the phase we wish to calculate. In general $\eta$ has both real and imaginary parts,
$\eta = \Re(\eta) + i \Im(\eta)$.
 Normalization implies that
\begin{equation}\label{norm}
1 =  \langle \psi(T) |\psi(T) \rangle =  e^{-2\Im(\eta)(T)} + \langle \psi_\bot (T) |\psi_\bot (T)\rangle 
\end{equation}
and  as time increases eventually the final state becomes $\psi_\bot (T)$.  The efficaciousness of the mode-invisibility method is in ensuring  that $\Im(\eta(T)) <<1$ for the time the probe is in the cavity.   

We work   perturbatively in the coupling strengths $(\lambda_p,\lambda_q)$ and extend our calculation of the evolution operator $U(0,T)$ to  second order. The phase factor $\eta$ in equation \eqref{phased1} has information about the entangled state \eqref{joint}.  Our task therefore is to find a way to measure this phase value provided that the criteria \eqref{hypo} and \eqref{phasee} are satisfied.

\subsection{Transition Probability}
 We consider now the effect  of the interaction between the probe and the entangled state. The probe
is initially in the state   $\ket{g_{p}}$ with constant speed $v$.   We calculate the probability that it gets excited after the interaction time $T=L/v$. If the entangled state is not perturbed, we expect that this excitation transition probability is approximately zero. x

To compute this probability, we note that the evolution of the system is, to second order in the coupling constant, given by
\begin{align}\label{rhot}
\rho(t) = \rho+ U^{(1)}\rho + \rho U^{(1)\dagger} + U^{(2)} \rho+\rho U^{(2)\dagger}+ U^{(1)}\rho U^{(1)\dagger}
\end{align}
 where $\rho(0)$ is the initial density operator for the combined state \eqref{joint} and $U^{(1)}, ~ U^{(2)}$ are the first and second order contributions in $(\lambda_p,\lambda_q)$ to the unitary operator. 
By tracing over the entangled state (ES), the only term contributing to the excitation probability is $U^{(1)} \rho(0) U^{(1)\dagger}$ so that after the interaction time $t>0$, the transition probability of exciting this detector is
\begin{align}\label{probb}
P_{  \ket{e_{p}}} = \langle e_p | \operatorname{Tr}_{\text{ES}}\Big[U^{(1)} \rho(0) U^{(1)\dagger} \Big] \ket{e_{p}}
\end{align} 
 $P_{  \ket{e_{p}}}$ depends on rotating wave terms and counter rotating wave terms coming from the Hamiltonian \eqref{hh}, with the rotating wave term yielding a higher contribution.  To ensure that we eliminate the order-$\lambda_{p}$ contribution to the excitation transition probability \eqref{probb}, we apply the mode-invisibility technique, obtaining the expression
\begin{align}\label{P_e}
P_{ \ket{e_{p}}}& =\lambda^{2}_{p} \Bigg[( |A|^{2}|\alpha|^{2}  + |B|^{2}|\beta|^{2}) | X_{+,\kappa}|^{2} +\sum_{\gamma} |X_{+,\gamma}|^{2}\Bigg]
\end{align}
for  the excitation transition probability $P_{ \ket{e_{p}}}$,
where the quantity $X_{+,\kappa}$ is defined in eq. (\ref{Xpm}) below and  $ X_{-,\kappa}$ vanishes  \cite{marvy2013, marvy2014}.  

Before proceeding, we note that the information gained from this approach is information about the initial state of the system.  Indeed, the qubit-cavity system evolves with time.  Provided $P_{ \ket{e_{p}}} \ll 1$  in \eqref{P_e}, the evolution
of the system is not disturbed by the probe.  Given the Hamiltonian for the system, the information gained from mode invisibility about the initial state is sufficient to reconstruct the evolution of the system at any subsequent time.

Furthermore, although we shall ensure that $P_{ \ket{e_{p}}} \ll 1$ for all of our parameter choices, we shall find that the phase $\eta$ in \eqref{phased1} is not small.  We proceed to compute $\eta$ in the next subsection.

 \begin{figure*}[t]
\includegraphics[width=.33\textwidth]{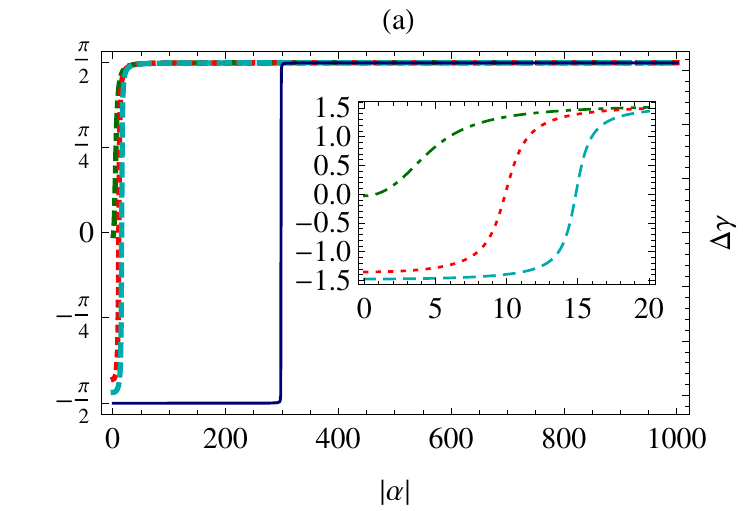} \includegraphics[width=.33\textwidth]{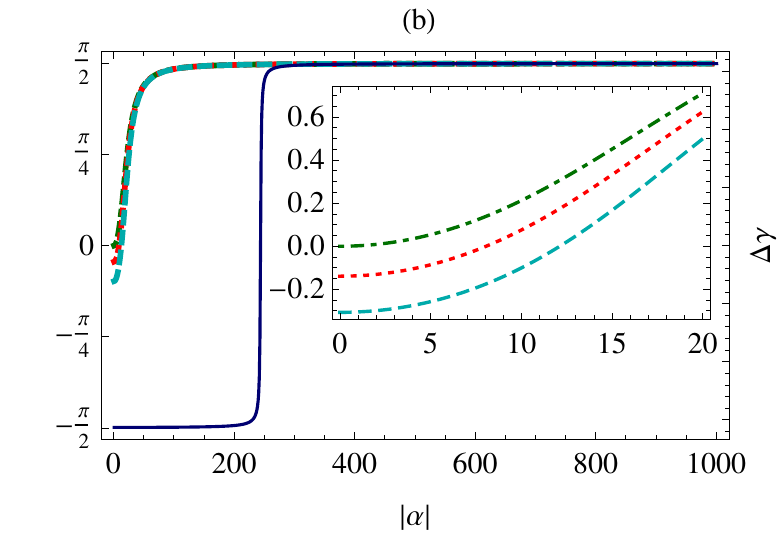}\includegraphics[width=.33\textwidth]{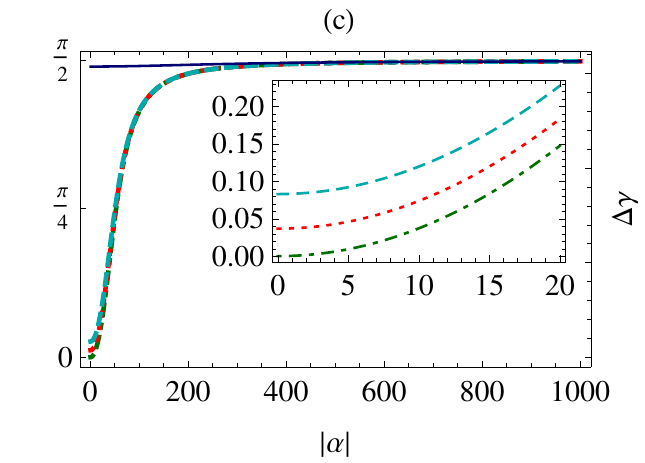}
\caption{(Color online): The phase factor $\Delta\gamma$ (defined in (\ref{rp}))  as a function of $|\alpha|$ for   $\theta = \pi/2$ and $\phi = -\pi/2$.
 Here, the qubit-field system is maximally entangled with $A=B=1/\sqrt{2}$. Different lines show  different values of $|\beta|$: $|\beta|=1$(green dotdashed Line), $|\beta|=10$ (red dotted  line), $|\beta|=15$ (Cyan dashed line), $|\beta|=300$(blue solid line). Inset shows this relation for a small range of $|\alpha|$ parameter. We compared different ratios of $\lambda_{q} =  r \lambda_{p}$: (a) $ r = 5$, (b) $ r= 1$, and (c) $r=10^{-2}$ and take $v = 10^{3}m/s$, $x_{0} = L/4$ where  $L \approx  0.019m$.  }
\label{onep}
\end{figure*}

\medskip
\subsection{Estimating the phase imprint of the probe}\label{phase}

 We now   compute  the phase acquired by the atom after its interaction time $T$ with the cavity mode. From equation \eqref{phased1}, this is given to leading order in both couplings as
\begin{align}\label{etadef}
\eta & =  -\ii \operatorname{ln}\Big[ \eta_{0} + \eta_{1} + \eta_{2}\Big] \nonumber\\
&=- \ii \operatorname{ln}\Big[1+ \langle \psi(0)| U^{(1)}|{\psi(0) }\rangle+ \langle \psi(0)| U^{(2)}|{\psi(0) }\rangle  \Big]
\end{align}
where $\operatorname{ln}$ is natural logarithm,  $U^{(1)}$ and  $U^{2}$ are the first and second order contributions to the evolution operator $U(0,T)$.  For convenience (and future use) we define
\begin{widetext}
\begin{align}
I_{\pm, \kappa} \equiv  \int_{0}^{T} \operatorname{dt} e^{\ii(\pm \Omega_{q} + \omega_{\kappa})t}\frac{\sin[k_{\kappa}x_0]}{\sqrt{\kappa \pi}}
\qquad
X_{\pm,\kappa} = \int_{0}^{T} \operatorname{dt} e^{\ii(\pm \Omega_{p} + \omega_{\kappa})t} \frac{\sin[k_{\kappa}vt]}{\sqrt{\kappa \pi}}
\label{Xpm}
\end{align}
\begin{align}\label{circprod}
 I_{\pm, \kappa}\circ I_{\pm, \delta}  \equiv
   \int_{0}^{T} \operatorname{dt_2}e^{\ii(\pm \Omega_{q} + \omega_{\kappa})t_2} \int_{0}^{t_2} \operatorname{dt_1}e^{\ii(\pm \Omega_{q} + \omega_{\delta})t_1} \frac{\sin[k_{\kappa}x_0]\sin[k_{\delta}x_0]}{\sqrt{\kappa \pi}\sqrt{\delta \pi}}
\end{align}
\begin{align}
 X_{\pm, \kappa}\circ X_{\pm, \delta}  \equiv
  \int_{0}^{T} \operatorname{dt_2}e^{\ii(\pm \Omega_{p} + \omega_{\kappa})t_2}
  \frac{\sin[k_{\kappa} vt_2]}{\sqrt{\delta \pi}}
 \int_{0}^{t_2}\operatorname{dt_1} e^{\ii(\pm \Omega_{p} + \omega_{\delta})t_1}    \frac{\sin[k_{\delta}vt_1]}{\sqrt{\delta \pi}}.
\end{align}
\end{widetext}
We obtain
\begin{align}\label{phase1}
\eta_{1} & =   -  \frac{2\ii \lambda_{q} }{\sqrt{k_{\kappa}L}} \Big(\operatorname{Re}\big[\big(I_{+\kappa}^{*}\beta + I_{-\kappa}\alpha^{*}\big) \langle \alpha\ket{\beta} A^{*} B\big]\Big)
\end{align}
to first order in $ \lambda_{q}$.

In general $\eta_1 \neq 0$.  It is purely imaginary and so must remain small to ensure that the contribution of
$\psi_\bot (T)$ in \eqref{norm} remains negligible.  There are several ways of doing this.  One is to ensure that the qubit
sits at a node so that $\sin[k_{\kappa}x_0] = 0$.  This will cause all $I_{\pm, \kappa}$ integrals to vanish,
and the results will be the same  as those obtained  for $B=0$ \cite{marvy2014}.  Another is to fine-tune the speed of
the probe (i.e. $v/c$) relative to the qubit phase so that $I_{\pm, \kappa}$ vanishes but $I_{\pm, \kappa} \circ I_{\pm, \kappa}^{*}$ will not.
A third approach is to choose the relative phase of the coherent states so that
  $\langle \beta \ket{\alpha} = e^{-\frac{1}{2}(|\beta|^{2} + |\alpha|^{2} - 2 \beta^{*}\alpha) }
= e^{-\frac{1}{2}(|\beta|+ |\alpha|)^{2} }$. Our choice of phase ensures this relation.  The maximal value of this quantity
is less than unity, and   over the range of  parameters we choose we find that $\Im(\eta) \leq 10^{-5}$ for all
$|\beta|$ and $|\alpha|$; for most of this range  it is many orders of magnitude smaller than this.

Since $\eta_1 $ does not depend on any parameters of the probe,   we  require $\eta_2$ in order
for the probe to be a useful diagnostic of the generalized CAT state.
\begin{figure*}[t]
\includegraphics[width=.33\textwidth]{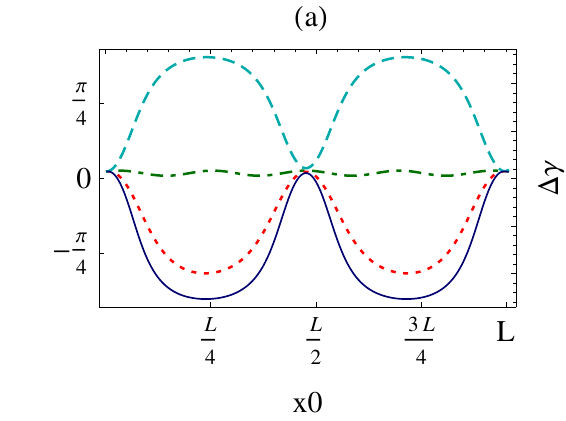} \includegraphics[width=.33\textwidth]{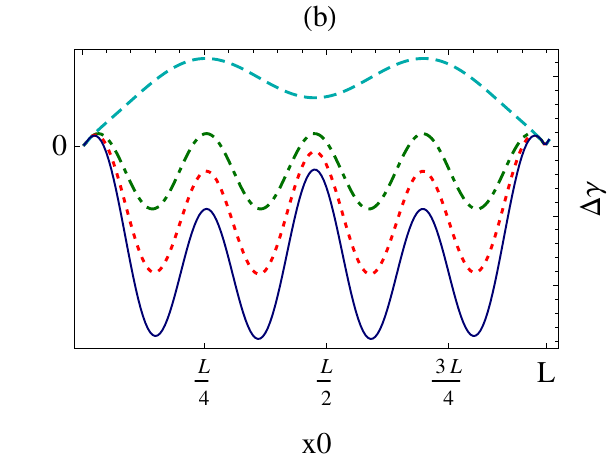}\includegraphics[width=.33\textwidth]{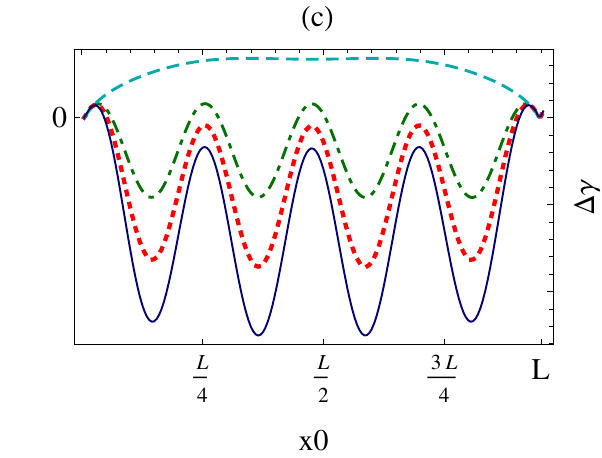}
\caption{(Color online):  The phase factor $\Delta\gamma$ (defined in (\ref{rp})) measured in Eq. \eqref{phased1} as a function of $x_{0}$-- the qubit's position  in the cavity.  Each graph plots different  values of $|\alpha| = |\beta|$ : (a) $ |\alpha| = |\beta|=10$, (b) $|\alpha|=|\beta|=1$ and (c) $|\alpha| = |\beta|=0.3$. The different lines within each graph illustrate  different values of $A,B$ : $A=1/2, B=\sqrt{3}/2$ (red  Line), $A=1/\sqrt{2}, B=1/\sqrt{2}$ (green  line), $A=1, B=0$ (cyan  line), $A=0, B=1$(blue line). Here we considered $\lambda_{q} = 3 \lambda_{p}$.    }
\label{qposition}
\end{figure*}
We obtain
\begin{widetext}
\begin{align}\nonumber
\eta_{2}&=  -\lambda_{q}^{2}\Big( I_{-\kappa}^{*} \circ I_{-,\kappa} B^{2}|\beta|^{2} +   I_{-,\kappa} \circ  I_{-,\kappa}^{*} A^{2} |\alpha|^{2}+ I_{+\kappa}\circ I_{+,\kappa}^{*} B^{2}|\beta|^{2} +   I_{+,\kappa}^{*} \circ  I_{+,\kappa} A^{2}|\alpha|^{2} + I_{-,\kappa}\circ I_{+,\kappa} (\alpha^{*})^{2} A^{2}+
\label{etato}
\\
&\quad  I^{*}_{-,\kappa}\circ I_{+,\kappa}^{*} B^{2} \beta^{2} + I_{+,\kappa}\circ I_{-,\kappa}B^{2}(\beta^{*})^{2}+ I_{+,\kappa}^{*} \circ I_{-,\kappa}^{*}A^{2} \alpha^{2} + \sum_{\delta}A^{2}I_{+,\delta}^{*}\circ I_{+,\delta}  + \sum_{\delta}B^{2} I_{-,\delta}^{*}\circ I_{-,\delta} \Big)-\\\nonumber
& \lambda_{p}^{2}\Big( X_{+\kappa}^{*}\circ X_{+,\kappa}  \big(A^{2} |\alpha|^{2} +B^{2}|\beta|^{2}\big) + X_{-,\kappa}X_{+,\kappa}\Big((\alpha)^{*} A^{2} + (\beta^{*})^{2}B^{2} \Big) + X_{+,\kappa}^{*}X_{-,\kappa}^{*}\Big( \alpha^{2}A^{2} + \beta^{2} B^{2}\Big)+  \sum_{\delta}X_{+,\delta}^{*}\circ X_{+,\delta}  \Big)
\end{align}
\end{widetext}
Note that  for a maximally entangled state $A=B=1/\sqrt{2}$ the terms on the first line above vanish for   $\vert\beta\vert = \vert\alpha\vert$.

\begin{figure*}[t]
 \includegraphics[width=.33\textwidth]{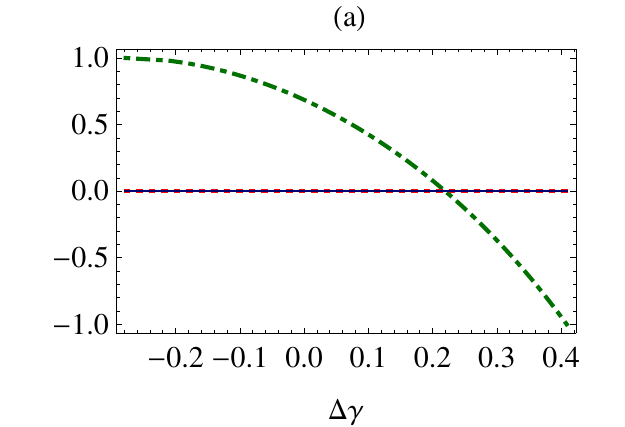}\includegraphics[width=.34\textwidth]{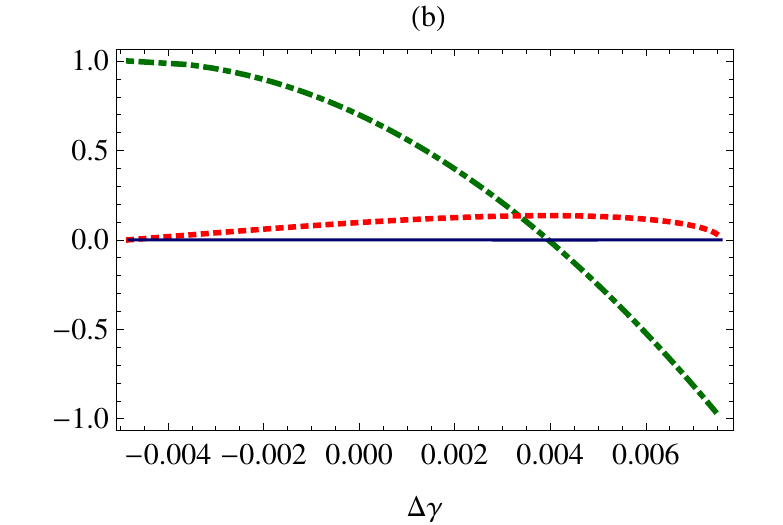}\includegraphics[width=.34\textwidth]{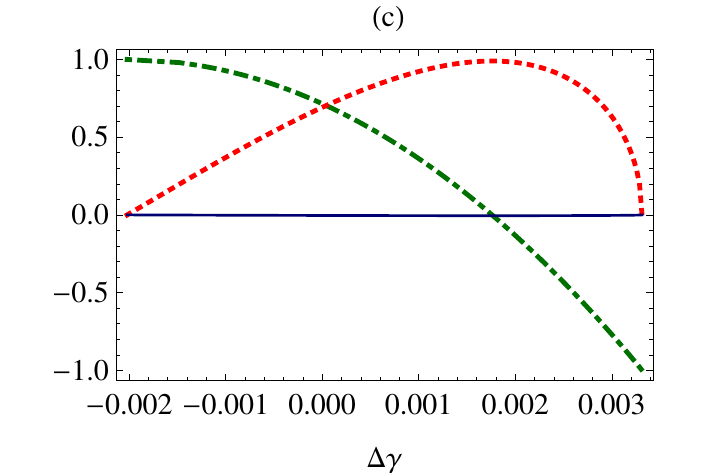}
\caption{ (Color online):   Population inversion (Green dotdashed line), dipole current (blue solid line) and dipole moment (red dotted line) as a function of interferometric phase difference $\Delta \gamma$ for different field amplitudes: (a) $|\alpha|=|\beta| = 10$, (b) $|\alpha|=|\beta| = 1$ (c) $|\alpha|=|\beta| = 0.1$ respectively. Here we considered the ratio $\lambda_{q} = \lambda_{p}$. With the normalization condition in mind $A^{2} + B^{2} = 1$, we start with $A=0, B=1$ and gradually increase $A$ up to 1.  }
\label{atominv}
\end{figure*}
\section{Measurement Process}\label{three}
 In section \ref{phase}, we calculated the phase $\gamma = \operatorname{\Re}[\eta]$ imparted on the probe   after an interaction time $T$. Since $\gamma$ is a global phase, the only way to measure it is via an interferometric setup \cite{marvy2013,marvy2014}. For this setup, a reference state is required which is contingent to the purpose of the experiment. For simplicity, suppose we have the vacuum state $\ket{0}$  in the reference cavity so that the joint initial state for this cavity is $\ket{g_p}\otimes \ket{0}$ where the state $\ket{g_p}$ is the state of the detector. In this cavity, the detector acquires a phase $\eta_{R}$ when it interacts with the vacuum state. After an interaction time $T$  the atomic interferometric phase difference is observed to be
\begin{align}\label{phre}
\Delta \eta= \eta -  \eta_{ R}
\end{align}
Typically speaking, $\Delta \eta$ is a measurable quantity and it reveals information about the unknown field state.  Following the procedure in section \ref{phase} we can obtain an expression for $\eta_{R}$
\begin{align}\label{refphase}
\eta_{R} =- \ii \operatorname{ln}[1- \lambda_{p}^{2}\sum_{\delta}X_{+,\delta}X_{+, \delta}^{*}]
\end{align}
which depends only on the vacuum terms. Therefore the interferometric  phase difference on the probe after the interaction time with the entangled state is
\begin{align}\label{rp}
\Delta \gamma = \Re[\Delta \eta]
\end{align}

In the following sections we will investigate the behaviour of the entangled state by carefully studying the relative phase factor \eqref{rp} within the coupling range $10^{-2} \leq \lambda_{q}/\lambda_{p} \leq 5$.

\subsection{Measurement of amplitude of the cavity field}

  In figure \ref{onep}, we plot the phase $\Delta\gamma$ from (\ref{rp}) as a function of $|\alpha|$ (the amplitude of the coherent state $\ket{\alpha}$). We see that $\Delta\gamma$  monotonically increases from $-\pi/2$ to $+\pi/2$ as $|\alpha|$ increases  (see Fig \ref{onep}), with the asymptotic value of $\pi/2$  effectively reached once $|\alpha|$ becomes sufficiently large. For the special case $\frac{\lambda_{q}}{\lambda_{p}} \ll 1$, one would expect that the probe-field interaction dominates and the entangled state behaves as if it were a coherent state in the cavity.  We indeed find this to be the case when we compare figure \ref{onep}(c) with the behaviour of a coherent state in a cavity probed by a qubit as considered in \cite{marvy2014}.  Note that $\Delta\gamma$ is most efficacious as a diagnostic for the value of $|\alpha|$ when $\lambda_{q} \ge \lambda_{p}$, as Fig \ref{onep}(a) and \ref{onep}(b) demonstrate: for a broad range of values of $|\beta|$ a measurement of $\Delta\gamma$ yields information about the value of $|\alpha|$.  However this becomes increasingly less so once $|\beta|$ becomes sufficiently large; as shown in Fig \ref{onep}(a), we see that  $\Delta\gamma$ is insensitive to the value of $\alpha$ except near  
  $|\beta| = |\alpha| = 300$.  In a narrow region near $|\beta| = |\alpha| = 300$, we observe the sudden rise of $\Delta\gamma$ to $\pi/2$, after which we lose information about the state amplitude $|\alpha|$. This sudden response deviates slightly to the left as $\lambda_{q} = \lambda_{p}$ (see Fig \ref{onep}(b)) until we have a zero response to $|\alpha|$ (Fig \ref{onep}(c)). 

\subsection{Measurement of the Cavity Qubit's Position}

Here we demonstrate that the mode invisibility technique can be used to  obtain information about the qubit's position in a cavity. Recall that we assumed that the transition frequency of the qubit is detuned from the cavity field frequency. 
 The probe's interaction with the entangled qubit-field system  imparts a phase shift $\Delta\gamma$ to the entangled state. The magnitude of this phase shift depends also on the qubit's position $x_{0}$ relative to the nodes and antinodes of the cavity modes. 

We illustrate the relationship between the interferometric phase difference as a function of qubit's position $x_{0}$ in  Fig \ref{qposition}.  The phase $\Delta\gamma$ oscillates  between maximal and minimal values that depend on the relative magnitude of $|\alpha|/ |\beta|$.  The amplitude of oscillation varies as the quantum state \eqref{estate} goes from being one of maximal entanglement to a product state. In the latter situation, for large $|\alpha| = |\beta|$ (Fig \ref{qposition}(a))
the qubit's position for $A = 0,B=1$ is completely out of phase relative to the case $A =1,B=0$; with equal but opposite amplitudes they cancel each other. The maximally entangled case $A = B=\frac{1}{\sqrt{2}}$ has near-zero amplitude.  

The qubit's position  (up to a wavelength) can therefore be determined from the phase, providing some   advantage for preparing the qubit in an entangled state. As $|\alpha|=|\beta|$ decreases in value the distinctions between the
two possible product states become increasingly pronounced.  This asymmetry is due to 
the distinction between  the less rapidly oscillating counter-rotating ($A$-coefficient) and more rapidly oscillating rotating ($B$-coefficient) vacuum contributions to $\eta_2$, which  dominate for small  $|\alpha|,  |\beta|$.  The increase in phase  oscilliation frequency makes it somewhat more difficult to determine the location of the qubit.  The phase symmetry for large $|\alpha|,  |\beta|$ becomes increasingly less valid, with large changes in
the phase occuring for $A = 0,B=1$ corresponding to small changes for  $A =1,B=0$ and vice-versa, as shown in Fig \ref{qposition}(c).

\subsection{Understanding the qubit's dynamics}\label{AT}


We close this section by exploring what properties of the state of the cavity qubit can be obtained using the mode invisibility technique. The  reduced density operator for the qubit is given by tracing over the detector and cat field variables:
\begin{align}\label{rhoq}\nonumber
\rho_{q}(t)=&\operatorname{Tr_{FP}}\Big[\rho(t)\Big]\\
=&\rho_{gg} |g\rangle\langle g| + \rho_{ge}|g\rangle \langle e| + \rho_{eg} |e\rangle \langle g| + \rho_{ee}|e\rangle \langle e|
\end{align}
where $\rho_{gg}, \rho_{ge}, \rho_{eg}$ and $\rho_{ee}$ are respectively defined in the appendix with the diagonal elements satisfying $\rho_{ee} + \rho_{gg}=1$ and the off-diagonal elements are in general complex and satisfy $\rho_{eg}=\rho_{ge}^{*}$. Eq. \eqref{rhoq} enables us to study some properties of the qubit.
\begin{figure*}[b]
\includegraphics[width=.33\textwidth]{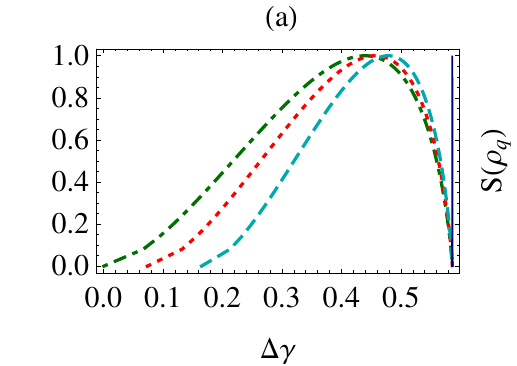} \includegraphics[width=.33\textwidth]{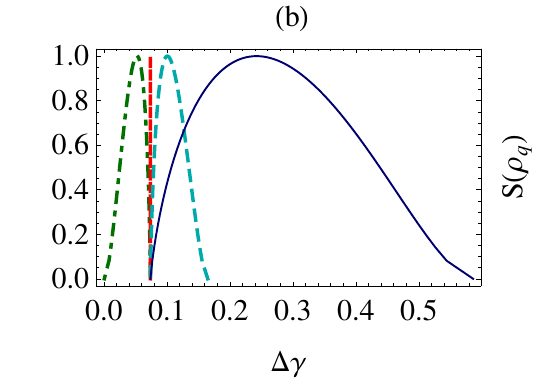}\includegraphics[width=.33\textwidth]{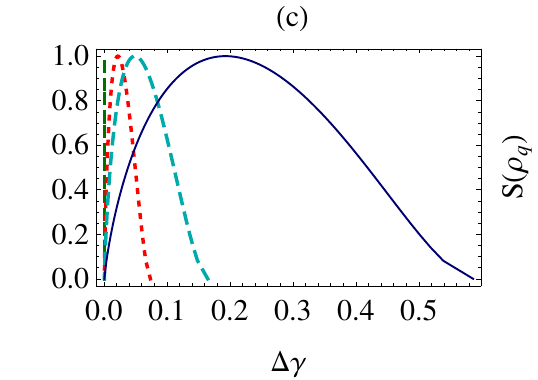}
\caption{(Color online): von Neumann entropy as a function of $\Delta\gamma$ for different $|\alpha|$; (a) $|\alpha|=30$ (b) $|\alpha| = 10$ (c) $|\alpha|=1$. Here $\lambda_{q}= 10^{-2} \lambda_{p}$. Different curves show different values of $|\beta|$: $|\beta|=1$ (green dotdashed line), $|\beta|=10$ (red dotted line), $|\beta|= 15$ (cyan dashed line), $|\beta|=30$ (blue solid line). }
\label{entropys1}
\end{figure*}

\begin{figure*}[t]
\includegraphics[width=.33\textwidth]{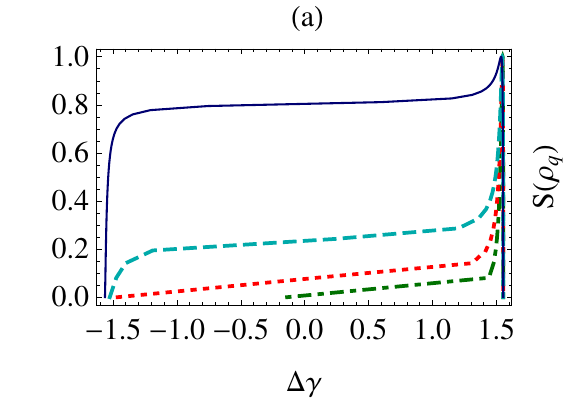} \includegraphics[width=.33\textwidth]{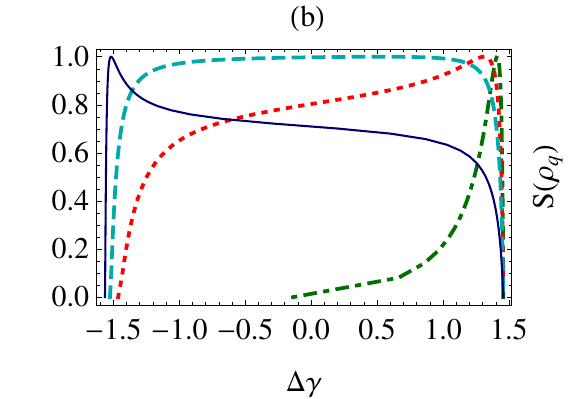}\includegraphics[width=.33\textwidth]{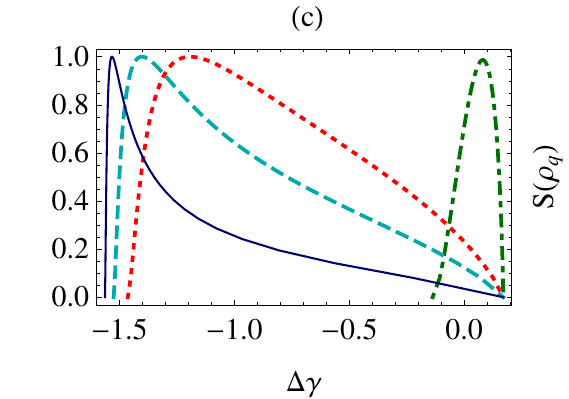}
\caption{(Color online): von Neumann entropy as a function of $\Delta\gamma$ for different $|\alpha|$; (a) $|\alpha|=30$ (b) $|\alpha| = 10$ (c) $|\alpha|=1$. Here $\lambda_{q}=5\lambda_{p}$. Different curves reveal different values of $|\beta|$: $|\beta|=1$ (green dotdashed line), $|\beta|=10$ (red dotted line), $|\beta|= 15$ (cyan dashed line), $|\beta|=30$ (blue solid line). 
}
\label{entropys2}
\end{figure*}
\begin{figure*}[t]
\includegraphics[width=.33\textwidth]{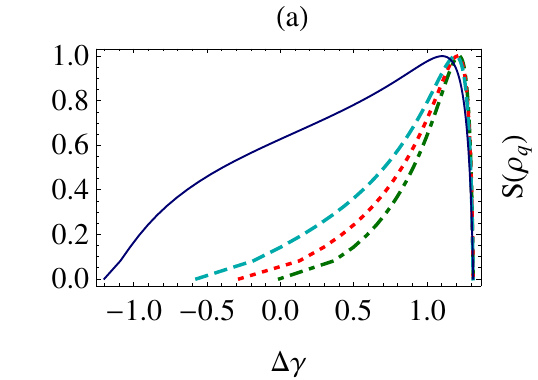} \includegraphics[width=.33\textwidth]{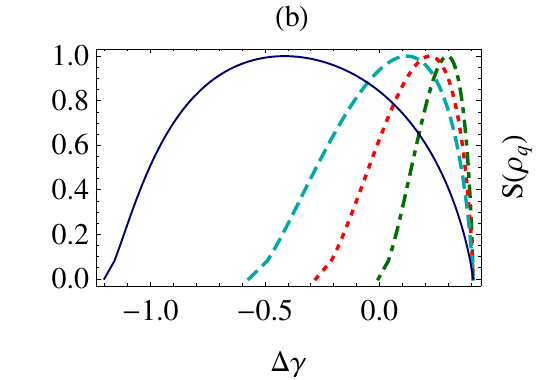}\includegraphics[width=.33\textwidth]{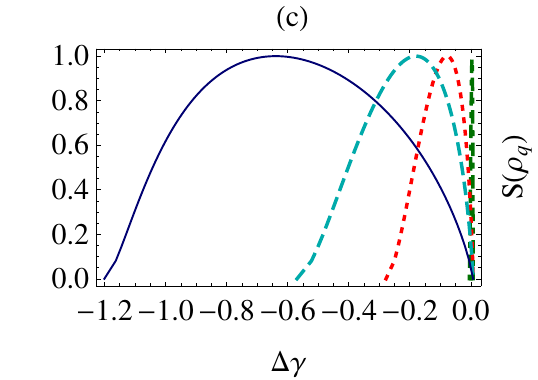}
\caption{(Color online) : von Neumann entropy as a function of $\Delta\gamma$ for different $|\alpha|$; (a) $|\alpha|=30$ (b) $|\alpha| = 10$ (c) $|\alpha|=1$. Here $\lambda_{q}= \lambda_{p}$. Different curves reveal different values of $|\beta|$: $|\beta|=1$ (green dotdashed line), $|\beta|=10$ (red dotted line), $|\beta|= 15$ (cyan dashed line), $|\beta|=30$ (blue solid line). }
\label{entropys3}
\end{figure*}

 We define the functions
\begin{align*}
A_{I}& = \rho_{ee} + \rho_{gg}   &\text{Probability}\\
A_{z}&=\rho_{ee}-\rho_{gg} =  \langle \sigma_{z}\rangle &\text{Population inversion}\\
A_{x}& = \rho_{eg} + \rho_{ge} = \operatorname{\Re}\lbrace \langle \sigma^{+}\rangle \rbrace   &\text{Dipole moment}\\
A_{y} &= \frac{1}{\ii} (\rho_{eg} - \rho_{ge}) =  \operatorname{\Im}\lbrace \langle \sigma^{+}\rangle \rbrace &\text{Dipole current}
\end{align*}
We plot in Fig. \ref{atominv} these functions against the interferometric phase difference $\Delta \gamma$.  Starting with $B=1$ and $A=0$, as  we increase $A$ gradually we see that the interferometric phase difference increases as the population inversion decreases. In figure \ref{atominv}(a), we see that for large values of $\alpha, \beta$, we see a zero dependence on $A_{x}$ and $A_{y}$. However while we increase the values of $|\alpha|,|\beta|$, although a response of $A_{y}$ is never seen, we see $A_{x}$ responds attaining a maximum at some intermediate positive value of $\Delta \gamma$. Within this limitation the phase difference provides a diagnostic of the cavity qubit's state.

 \section{Characterizing the entanglement}\label{four}

\begin{figure*}[t]
\includegraphics[width=.33\textwidth]{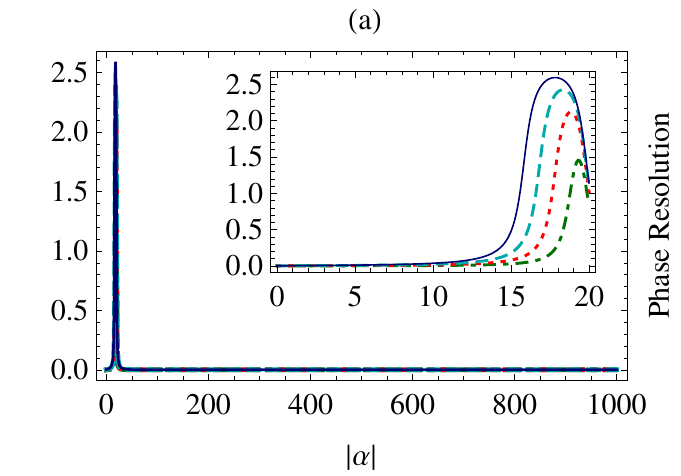} \includegraphics[width=.33\textwidth]{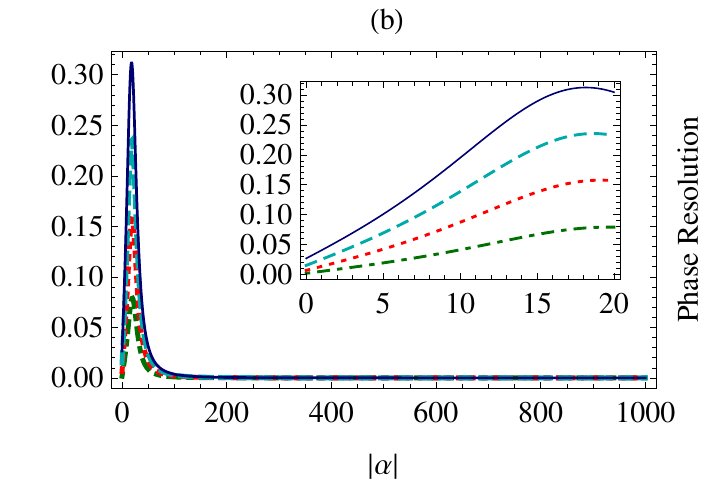}\includegraphics[width=.33\textwidth]{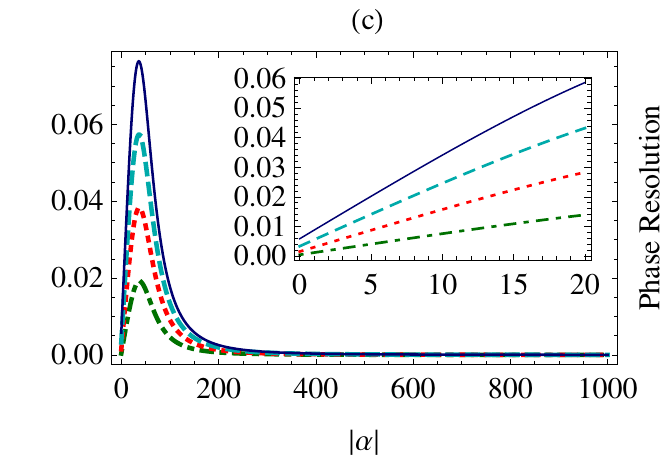}
\caption{(Color online): The phase resolution $R_{\delta\alpha}(\alpha)$ as a function of $|\alpha|$ for different ratios $\lambda_{q} = r\lambda_{p}$, (a) $r = 5$, (b) $r = 1$ and (c) $r=10^{-2}$. In this case, the qubit is fixed at $x_0=L/4$ and we use the values $\theta=\pi/2$,$\phi=-\pi/2$, $|\beta|=20$ and different values of $\delta\alpha$: $\delta\alpha=1$ (green dotdashed line), $\delta\alpha=2$ (red dotted line),$\delta\alpha=3$ (cyan dashed line), $\delta\alpha=4$ (blue solid line). } 
\label{phaseresoff}
\end{figure*}

\begin{figure*}[t]
\includegraphics[width=.33\textwidth]{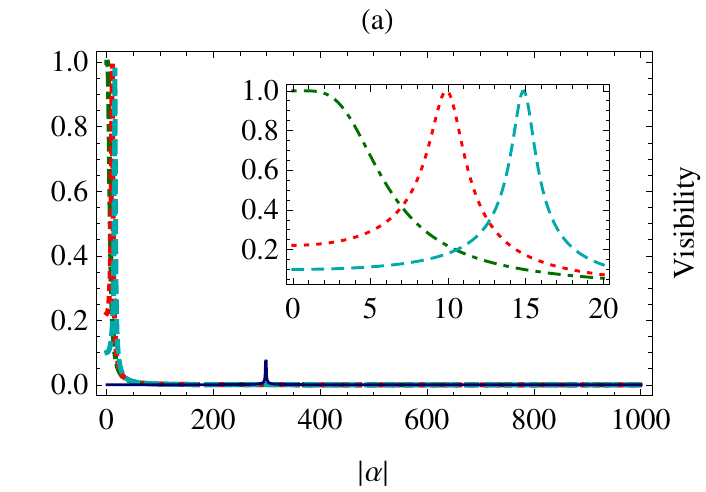} \includegraphics[width=.33\textwidth]{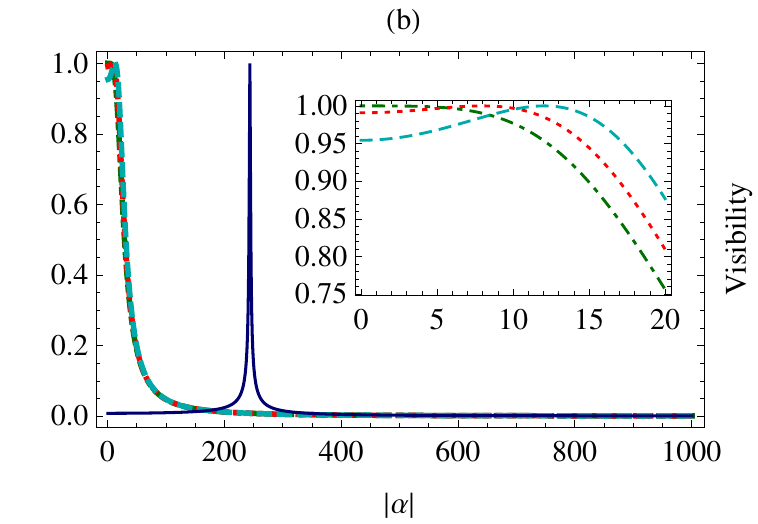}\includegraphics[width=.33\textwidth]{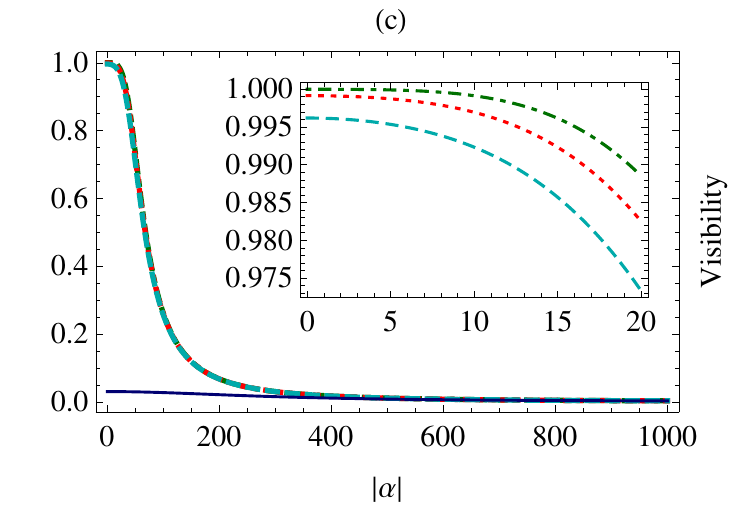}
\caption{(Color online):Visibility $\text{Exp}[\Im[\eta]]$ as a function of $|\alpha|$ . The qubit is fixed at $x_0=L/4$ and we use the values $\theta=\pi/2$,$\phi=-\pi/2$ and different values of $|\beta|$: $|\beta|=1$ (green dotdashed line),$|\beta|=10$ (red dotted line), $|\beta|=15$ (cyan dashed line),$|\beta|=300$ (blue solid line). We see a peak for the case $\lambda_{q} = 5 \lambda_{p}$. This occurs at point $|\alpha|=|\beta|$. This peak gradually shifts from this position as $\lambda_{q}$ becomes smaller compared to $\lambda_{p}$ until it dies out. } 
\label{visi}
\end{figure*}

  The previous section demonstrated that an entangled state of the form \eqref{estate} 
can be non-destructively probed under certain conditions, the probe acquiring a phase factor \eqref{etadef}. This phase is measurable and carries information about the  entangled state, as discussed in section \ref{three}, where we assumed that we have no knowledge of the physical features of the composite of the entangled state however we have knowledge of the degree of entanglement $A$ and $B$.

In this section, we  consider another possibility for  probing states of the form in Eq. \eqref{estate}: we assume we have knowledge of the features of the entangled state and we want to measure the entanglement between the field and the qubit after the interaction.

Although several entanglement measures exist \cite{Bennett1895,Bennett1996,Bennett19962,Emeasures,QM}, we choose the von Neumann entropy \cite{QIP} of the reduced state $S(\rho_{q})$ which is given as
\begin{align}\label{Erho}
 S(\pi_{i} ) = \begin{cases} -\sum_{i}  \pi_{i} \operatorname{Log} \pi_{i}
        & \quad \text{if } \pi_{i} > 0\\
  0 & \quad \text{if }\pi_{i} = 0\\
  \end{cases}
\end{align}
where $\rho_{q} = \rho_{q}(t)= \operatorname{Tr}_{\text{FP}}[\rho(t)] = \sum_{i} \pi_{i} |i \rangle \langle i|$  is the qubit's reduced density operator given in \eqref{rhoq}, with eigenvalues $\pi_{j}$ given as
\begin{align}\label{eigenvalues}
\pi_{\pm j} =  \frac{1}{2} \Big(\rho_{gg} +\rho_{ee} \pm 
 \sqrt{4 \rho_{ge}\rho_{eg}+ (\rho_{ee}-\rho_{gg})^{2}}\Big).
\end{align}
 Fixing the qubit  at $x_0=L/4$, we consider the values $\theta=\pi/2, ~ \phi=-\pi/2$.  We plot the resultant von Neumann entropy $S(\rho_{q})$ as a function of $\Delta\gamma$ for three distinct ratios of the couplings: $\lambda_{q} = r \lambda_{p}$ with $r=10^{-2}, 5,$ and $1$ in figures \ref{entropys1}, \ref{entropys2} and \ref{entropys3} respectively. We find that $\Delta\gamma$ provides an excellent measure of the entropy over a broad range of $|\alpha|$ and $|\beta|$ for all coupling ratios $\frac{\lambda_{q}}{\lambda_{p}}$ explored.  However for small $\lambda_q/\lambda_p$, the range of $\Delta\gamma$ becomes increasingly narrow as $|\alpha|/|\beta| \to 1$, becoming very sharply peaked in this limit. 

We find that the maxima  of $ S(\rho_{q})$ are less than unity for small values of $ |\alpha|,|\beta| \leq 1$. Setting $ |\alpha|=|\beta|$, we find that these maxima approach unity as $|\alpha|\to 1$   (see the green curves in Fig $5c$, $6c$ and $7c$), with little change for larger values of $|\alpha|$.   Alternatively if we fix  $|\alpha|$,  the maxima increase with increasing $|\beta|$ until attaining $  S(\rho_{q}) = 1$ (within our limits of numerical precision) after which they gradually decrease.  
 
In summary, the von Neumann entropy for the reduced qubit state (which is equal to the von Neumann entropy for the cavity cat state) increases with an increase in the cavity cat amplitudes $|\alpha|$ and $|\beta|$. This provides us with a tool for measuring the entropy of a quantum system after interaction, a question of common interest  \cite{QIP}.  Since the mode invisibility  process does not significantly affect the entanglement between the cavity qubit and the field,  we infer that $\Delta\gamma$ is providing a measure of the constant entropy between them.

\section{Phase Resolution and Interferometric Visibility}\label{six}

For any given setting, the  atom interferometer can resolve the values of the parameters only to a certain level of precision. The phase resolution of the interferometric setting is 
\begin{align}\label{res}
R_{\delta\alpha}(\alpha)=| \gamma_{\alpha+\delta\alpha} -  \gamma_{ \alpha}|
\end{align}
and by fixing a value for $|\beta|$, Eq. \eqref{res} defines our ability to distinguish the phase acquired by the state  having amplitude $|\alpha + \delta \alpha|$ from another of amplitude $|\alpha|$. We illustrate in figure \ref{phaseresoff} the phase resolution for a fixed value of $\vert\beta\vert=20$. 
We see that the phase discrimination is quite good, provided $|\alpha|$ is not too large, with the best discrimination over the broadest range of $|\alpha|$ occurring for $ {\lambda_{q}} = {\lambda_{p}}$.

 We next consider the visibility of the interferometric fringes.  As discussed in section \ref{mc}, $\eta$ is a complex number due to   contributions from terms orthogonal to our initial chosen quantum state. This yields an observable loss in the visibility, where
\begin{align}
|\langle \Psi (T) \ket{\Psi(0)}|^2 =   \exp[-2\Im[\eta]] 
\end{align}
defines the visibility, with $\ket{\Psi(T)}$  the final quantum state at time $T$. This provides a measure of how non-destructive the probe is.
We present in figure \ref{visi}  the visibility factor as a function of $|\alpha|$ for different regimes of $\lambda_{q}$ and $\lambda_{p}$.   

For small values of $|\beta|$ visibility remains largest over the broadest range of $|\alpha|$ for
$ {\lambda_{q}} = {\lambda_{p}}$;  $|\alpha|<100$ yields non-destructive measurement capacity for
the probe at better than 99\%.  There is not much decline in visibility over the same range of
$|\alpha|$ as $ {\lambda_{q}}$ decreases.  However as $ {\lambda_{q}}$ increases, visibility gets notably weaker over increasingly small ranges of $|\alpha|$. However we also see an interesting trend as $|\beta|$ increases:  for a given value of $|\beta|$
visibility peaks, with the location of the peak occurring approximately at $|\alpha|= |\beta|$.  At this point, we say the two coherent states $\ket{\alpha}, \ket{\beta}$ coincide.

\section{Conclusion}

 We have demonstrated  the utility of the mode-invisibility measurement technique \cite{marvy2013} for non-destructively probing an entangled generalized qubit/cat state in a cavity mode.  For realistic physical parameters, and provided that the amplitude of the cavity field is not too large, the technique works very well,  especially in the regime where $\lambda_q \approx \lambda_p$. However it breaks down once $|\alpha|$ is sufficiently large, which is where $\Delta\gamma$ reaches its asymptotic value of $\pi/2$. This is confirmed by means of the visibility, where we get up to $99\%$ non-demolition probe of the state for values of $|\alpha|$ close to $|\beta|$. In regimes where the mode-invisibility technique works,  the phase resolution is generally very good.
We also investigated the dynamics of the qubit state and the von Neumann entropy of the combined system. 
 Using the interferometric measurement setup \cite{marvy2013}, one could in principle be able to non destructively probe the entanglement of systems like \eqref{estate}.

 We close by noting that the suppression of  $\eta_1$ is sensitive to the relative choice of phase   $\theta=- \phi$ for
the coherent states.  By departing from these values In fact it is possible to modify our scheme to probe $(\theta, \phi)$ (up to an overall phase) for all possible values of $v$, as we shall demonstrate in future work.   It likewise should be possible  to explore the resonant regime  $ \omega_{\kappa} - \Omega_{q}=0$ where again $\eta_1$  is not suppressed. This induces an interaction between the qubit and probe, and so an extra term describing the probe-qubit interaction should be introduced  in the interaction Hamiltonian \eqref{hh}  to account for this.

\section{Acknowledgments}

This work was supported in part by the Natural Sciences and Engineering Research Council of Canada.  Paulina C-U was supported by CONACyT. Marvellous Onuma-Kalu thanks Kae Nemato and her group at the National Institute of Information, Tokyo, Japan for valuable comments.  We are grateful to Eduardo Martin-Martinez for a helpful discussion.  Both M. Onuma-Kalu and P. Corona-Ugalde contributed equally to this work.

\begin{widetext}

\appendix

\section{The Reduced Qubit State}

The density matrix for the reduced qubit state is defined by tracing out the cat field and detector variables:
\begin{align*}
\rho_{q}(t) =&\operatorname{Tr_{FP}}\Big[\rho^{(0)}+ U^{(1)}\rho + \rho U^{(1)\dagger} + U^{(2)} \rho+\rho U^{(2)\dagger}+ U^{(1)}\rho U^{(1)\dagger}\Big]\\
=&\rho_{gg} |g\rangle\langle g| + \rho_{ge}|g\rangle \langle e| + \rho_{eg} |e\rangle \langle g| + \rho_{ee}|e\rangle \langle e|
\end{align*}
Denoting
\begin{align*}                                                                                                                                                                                                                                                                                                                                                                                                                                                                                                                                                                                                                           \lbrace I_{+,\kappa}, I_{-,\kappa}^{*}\rbrace = I_{+,\kappa} \circ I_{-,\kappa}^{*}  + I_{-,\kappa}^{*} \circ I_{+,\kappa}
\end{align*}
and similarly for other circle products (defined in \eqref{circprod}), we obtain 

\begin{align*}
\rho_{gg}=&A^{2}-A^{2}\lambda_{q}^{2}\Bigg[\lbrace I_{+,\kappa}^{*}, I_{+,\kappa} \rbrace |\alpha|^{2}+ \lbrace I_{-,\kappa}^{*}, I_{-,\kappa} \rbrace |\alpha|^{2}+\lbrace I_{-,\kappa},I_{+,\kappa}\rbrace ( \alpha^{*})^{2}+ \lbrace I_{+,\kappa}^{*},I_{-,\kappa}^{*}\rbrace (\alpha)^{2} + \sum_{\gamma}\lbrace I_{+,\gamma}^{*},I_{+,\gamma}\rbrace\Bigg]\\
&-A^{2}\lambda_{p}^{2}\Big(\lbrace X_{+,\kappa}^{*}, X_{+,\kappa} \rbrace |\alpha|^{2}+ \lbrace X_{-,\kappa}^{*}, X_{-,\kappa} \rbrace |\alpha|^{2}+\lbrace X_{-,\kappa}, X_{+,\kappa}\rbrace ( \alpha^{*})^{2}+ \lbrace X_{+,\kappa}^{*},X_{-,\kappa}^{*}\rbrace (\alpha)^{2} + \sum_{\gamma}\lbrace X_{+,\gamma}^{*},X_{+,\gamma}\rbrace\Big)\\
&+\lambda_{q}^{2}B^{2}\Big[|I_{+,\kappa}|^{2}|\beta|^{2}+ 2\operatorname{Re}[I_{-,\kappa}I_{+,\kappa}( \beta^{*})^{2}]+  \sum_{\gamma}|I_{-,\gamma}|^{2}+  | I_{-,\kappa}|^{2}|\beta|^{2} \Big]+\lambda_{p}^{2}B^{2}\Big[ \sum_{\delta}|X_{+,\delta}|^{2}  +| X_{+,\kappa}|^{2} |\alpha|^{2}\Big]\\
&-\ii \lambda_{q} AB\Big( I_{+,\kappa}^{*}\beta+ I_{-,\kappa}\alpha^{*}\Big)\langle \alpha \ket{\beta}  + \ii \lambda_{q}AB\Big( I_{+,\kappa}\beta^{*} + I_{-,\kappa}^{*}\alpha\Big) \langle \beta \ket{\alpha} \end{align*}
\begin{align*}
\rho_{ge} =&A B  \langle  \beta | \alpha  \rangle-\lambda{q}^{2} AB  \langle \beta | \alpha \rangle \Big(\lbrace I_{+,\kappa}^{*},I_{+,\kappa} \rbrace \alpha \beta^{*} + \lbrace I_{-,\kappa}^{*},I_{-,\kappa} \rbrace \alpha \beta^{*} + \lbrace I_{+,\kappa},I_{-,\kappa} \rbrace (\beta^{*})^{2}+ \lbrace I_{+,\kappa}^{*},I_{-,\kappa}^{*} \rbrace \alpha^{2}\Big)\\
&-\langle \beta | \alpha \rangle AB \Big(\lbrace X_{+,\kappa}^{*},X_{+,\kappa}\rbrace \beta^{*}\alpha + \sum_{\delta}\lbrace X_{+,\delta}^{*},X_{+,\delta}\rbrace + \lbrace  X_{+,\kappa}^{*},X_{-,\kappa}^{*}\rbrace \alpha^{2} +\lbrace X_{-,\kappa},X_{-,\kappa}^{*}\rbrace \beta^{*} \alpha + \lbrace X_{-,\kappa},X_{+,\kappa}\rbrace  (\beta^{*})^{2}\Big]+\\
&\Big[  I_{+,\kappa}^{*}I_{-,\kappa}^{*} \beta \alpha^{*}   + (I_{+,\kappa}^{*})^{2}\beta^{2}  +\sum_{\gamma}I_{-,\gamma}I_{+,\gamma}^{*}+I_{-,\kappa}I_{+,\kappa}^{*} \alpha^{*}\beta+ (I_{-,\kappa})^{2} (\alpha^{*})^{2}\Big]AB \langle \alpha_{\kappa} | \beta_{\kappa} \rangle+\\
&\lambda_{p}^{2} AB \langle \beta | \alpha \rangle\Big(|X_{+,\kappa}|^{2} \beta^{*} \alpha + \sum_{\delta}|X_{+,\delta}|^{2}  \Big) +B^{2}\lambda_{q}^{2}\Big[I_{+\kappa}^{*} \beta + I_{-,\kappa} \beta^{*}\Big] +  A^{2}\Big[ I_{+\kappa}^{*} \alpha+ I_{-,\kappa}\alpha^{*} \big] 
\end{align*}
\begin{align*}
\rho_{eg} =&A B  \langle \alpha | \beta \rangle-\lambda{q}^{2} AB  \langle \alpha | \beta \rangle \Big(\lbrace I_{+,\kappa}^{*},I_{+,\kappa} \rbrace \alpha^{*} \beta + \lbrace I_{-,\kappa}^{*},I_{-,\kappa} \rbrace \alpha^{*} \beta + \lbrace I_{+,\kappa},I_{-,\kappa} \rbrace (\alpha^{*})^{2}+ \lbrace I_{+,\kappa}^{*},I_{-,\kappa}^{*} \rbrace \beta^{2}\Big)\\
&-\langle \alpha | \beta \rangle AB \Big(\lbrace X_{+,\kappa}^{*},X_{+,\kappa}\rbrace \beta \alpha^{*} + \sum_{\delta}\lbrace X_{+,\delta}^{*},X_{+,\delta}\rbrace + \lbrace  X_{+,\kappa}^{*},X_{-,\kappa}^{*}\rbrace \beta^{2} +\lbrace X_{-,\kappa},X_{-,\kappa}^{*}\rbrace \beta \alpha^{*} + \lbrace X_{-,\kappa},X_{+,\kappa}\rbrace  (\alpha^{*})^{2}\Big]+\\
&\Big[  I_{+,\kappa}^{*}I_{-,\kappa}^{*} \beta^{*} \alpha   + (I_{+,\kappa})^{2}(\beta^{*})^{2}  +\sum_{\gamma}I_{-,\gamma}^{*}I_{+,\gamma}+I_{-,\kappa}^{*}I_{+,\kappa}\alpha \beta^{*}+ (I_{-,\kappa}^{*})^{2} (\alpha)^{2}\Big]AB \langle \beta_{\kappa} | \alpha_{\kappa} \rangle+\\
&\lambda_{p}^{2} AB \langle \beta | \alpha \rangle\Big(|X_{+,\kappa}|^{2} \beta^{*} \alpha + \sum_{\delta}|X_{+,\delta}|^{2}  \Big) - \ii A^{2}\lambda_{q} \Big[ I_{+\kappa} \alpha^{*}+ I_{-,\kappa}^{*}\alpha \big] + \ii B^{2}\lambda_{q}\Big[I_{+\kappa} \beta^{*} + I_{-,\kappa}^{*} \beta \Big] \end{align*}
\begin{align*}
\rho_{ee} =&B^{2}-B^{2}\lambda_{q}^{2}\Bigg[\lbrace I_{+,\kappa}^{*}, I_{+,\kappa} \rbrace |\beta|^{2}+ \lbrace I_{-,\kappa}^{*}, I_{-,\kappa} \rbrace |\beta|^{2}+\lbrace I_{-,\kappa},I_{+,\kappa}\rbrace ( \beta^{*})^{2}+ \lbrace I_{+,\kappa}^{*},I_{-,\kappa}^{*}\rbrace (\beta)^{2} + \sum_{\gamma}\lbrace I_{-,\gamma}^{*},I_{-,\gamma}\rbrace\Bigg]\\
&B^{2}\lambda_{p}^{2}\Big(\lbrace X_{+,\kappa}^{*}, X_{+,\kappa} \rbrace |\beta|^{2}+ \lbrace X_{-,\kappa}^{*}, X_{-,\kappa} \rbrace |\beta|^{2}+\lbrace X_{-,\kappa}, X_{+,\kappa}\rbrace ( \beta^{*})^{2}+ \lbrace X_{+,\kappa}^{*},X_{-,\kappa}^{*}\rbrace (\beta)^{2} + \sum_{\gamma}\lbrace X_{+,\gamma}^{*},X_{+,\gamma}\rbrace\Big)\\
&+\lambda_{q}^{2}A^{2}\Big[|I_{+,\kappa}|^{2}|\alpha|^{2}+ 2\operatorname{Re}[I_{-,\kappa}I_{+,\kappa}( \alpha^{*})^{2}]+  \sum_{\gamma}|I_{-,\gamma}|^{2}+  | I_{-,\kappa}|^{2}|\alpha|^{2} \Big]+\lambda_{p}^{2}B^{2}\Big[ \sum_{\delta}|X_{+,\delta}|^{2}  +| X_{+,\kappa}|^{2} |\alpha|^{2}\Big]\\
&-\ii \lambda_{q} AB\Big( I_{+,\kappa}\beta^{*}+ I_{-,\kappa}^{*}\alpha\Big)\langle \beta \ket{\alpha}  + \ii \lambda_{q}\Big( I_{+,\kappa}^{*}\beta+ I_{-,\kappa}\alpha^{*}\Big) \langle \alpha \ket{\beta} 
\end{align*}
\end{widetext}

\bibliography{references}

\end{document}